\documentclass[preprint]{revtex4}

\usepackage{graphicx,psfrag}

\begin{document}

\title{Smile dynamics -- a theory of the implied leverage effect}

\author{Stefano Ciliberti, Jean-Philippe Bouchaud, Marc Potters}
\affiliation{Science \& Finance, Capital Fund Management, 6 Bd Haussmann, 75009 Paris, France}

\date{\today}

\begin{abstract}
We study in details the skew of stock option smiles, which is induced by the so-called 
leverage effect on the underlying  -- i.e. the correlation between past returns and future square 
returns. This naturally explains the anomalous dependence of the skew as a function of maturity of 
the option. The market cap dependence of the leverage effect is analyzed using a one-factor model.
We show how this leverage correlation gives rise to a non-trivial smile dynamics, which turns
out to be intermediate between the ``sticky strike'' and the ``sticky delta'' rules. Finally, we compare our result 
with stock option data, and find that option markets overestimate the leverage effect by a large factor, in particular
for long dated options.
\end{abstract}

\maketitle


\section{Introduction\label{sec:introduction}}

It is by now well known that the Black-Scholes model is a gross oversimplification
of real price changes. Non Gaussian effects have to be factored in order to explain (or 
possibly to predict) the shape of the so-called volatility smile, i.e. the dependence of the
implied volatility of options on the strike price and maturity. Market prices of options, once converted
into an effective Black-Scholes volatility, indeed lead to volatility {\it surfaces} -- the
implied volatility depends both on the strike price $K$ and the maturity $T$. But matters are
even more complicated since the whole surface is itself time dependent: the implied volatility 
associated to a given strike and maturity changes from one day to the next (see e.g. \cite{Rama}). 
An adequate model for 
the dynamics of the volatility surface is crucial for volatility risk management and for
market making, for instance. Market practice on this issue follows simple rules of thumb \cite{Hull}, such 
as the ``sticky strike'' rule where the volatility of a given strike and maturity remains constant
as time evolves. Another rule is ``sticky delta'' (or ``sticky moneyness'') where the volatility 
associated to a given moneyness stays constant, which means, to a first approximation, that the 
volatility smile is anchored to the underlying asset price. 

From a theoretical point of view, different routes have been suggested over the years to handle 
smile dynamics. One is to rely on local volatility models, popularized by Dupire \cite{Dupire} and Derman and Kani \cite{Derman},
where the underlying is assumed to follow a (geometric) Brownian motion with a local value of 
the volatility that depends deterministically on the price level and time. As shown by Dupire, it is
always possible to choose this dependence such that the smile surface is fitted exactly. This has 
been considered by many to be a very desirable feature, and such an approach has had considerable 
success among quants. Unfortunately, this idea suffers from lethal drawbacks. For one thing, this approach
predicts that when the price of the underlying asset decreases, the smile shifts to higher prices 
and vice versa (see the detailed discussion of this point in \cite{SABR}). This is completely opposite to what is observed in reality, where 
asset prices and market smiles tend to move in the same direction. From a more fundamental point of view, 
local volatility models cannot possibly represent a plausible dynamics for the underlying. The whole
approach is, in our mind, a victory of the ``fit only'' approach to derivative pricing and the demise of {\it theory}, 
in the sense of a true, first principle understanding of option prices in terms of realistic models of asset prices. 
Contrarily to what many quants believe, even a perfect {\it fit} of the smile is not necessarily a good
{\it model} of the smile.

A more promising path starts from realistic models for the true dynamics of the underlying. Many models have been proposed in order
to account for the non-Gaussian nature of price changes: jumps and L\'evy processes \cite{RamaBook}, 
GARCH and stochastic volatility models \cite{Gatheral,PHL},
multiscale multifractal models \cite{Muzy,lisa}, mixed jumps/stochastic vol models, etc. One particularly elaborate model of 
this kind is the so-called SABR model, 
where the log-volatility follows a random walk correlated with the price itself in order to capture the leverage effect, i.e.
the rise of volatility when prices fall (and vice-versa, see below). Interestingly, all these models predict a certain term structure
for the cumulants (skewness, kurtosis) of the return distribution over different time scales, which in turn allows one to calculate 
the parabolic shape of the smile for near-the-money options of different maturities $T$ \cite{Backus,CPB,Book,SABR}. For example, models where returns 
are independent, 
identically distributed -- IID -- variables (like L\'evy processes), the skewness is predicted to decay like $T^{-1/2}$ and the kurtosis as $T^{-1}$.
As we show below, the leverage effect leads to a much richer
term structure of the skewness, whereas long-ranged volatility clustering leads to a non trivial term structure of the kurtosis \cite{Book}.

The aim of this paper is to explore the dynamics of the smile around the money, within the lowest order approximation that only retains 
the skewness effect. We show that such an approximation leads to an explicit prediction for dynamics of the smile, in particular of 
the {\it implied} leverage effect, i.e., the correlation between returns and at the money implied volatilities. 
Our result only depends on the historical leverage correlation, and predicts a volatility shift intermediate between ``sticky strike''
(for short maturities) and ``sticky delta'' (for long maturities), in a way that we detail below. We then compare our result with 
market option data on stocks and indices, and find that the market on average overestimates the leverage effect by a rather large factor.

\section{Volatility Smile: Cumulant Expansion and Historical Leverage}

Let us first recall the cumulant expansion of the volatility smile, worked out in slightly different form in several papers
(see \cite{Backus,CPB,Book} and also \cite{Sircar,SABR,Durrlemann}). Converted into a Black-Scholes volatility $\Sigma$, the price of a near-the-money
option of maturity $T$ can be generally expressed as:
\begin{equation}
  \Sigma(K,T) = \sigma \left[1 + \frac{\zeta(T)}{6} {\cal M} + \frac{\kappa(T)}{24} ({\cal M}^2-1) + {\cal O}({\cal M}^3)\right]
\end{equation}
where $K$ is the strike, $S$ the price of the underlying, $\sigma$ the true volatility of the stock and 
${\cal M} = (Ke^{-rT}/S-1)/\sigma\sqrt{T}$ is the \emph{moneyness}. $\zeta(T)$ and $\kappa(T)$ are respectively 
the skewness and the kurtosis of the forward looking, un-drifted probability of price changes over lag $T$ \cite{Book}. In the above formula
we neglect various terms that are usually small (for example, a term in $\zeta^2 {\cal M}^2$ that is small compared
to the kurtosis contribution). In the following we will in fact discard the quadratic contribution and study the assymmetry 
of the volatility smile for options in the immediate vicinity of the money, where the smile is 
entirely described by the volatility and the skewness. Although both these quantities should be interpreted as forward looking,
it proves very useful to see what a purely historical approach has to say. The unconditional historical skewness can be written 
in full generality as \cite{Book}:\footnote{In fact, this formula assumes that the three-point return cumulant, 
$\langle r_i r_j r_k \rangle_c$, is zero when $i \neq j \neq k$. We have checked empirically that this term is small, even summed 
over non coinciding times.}
\begin{equation}
  \zeta(T) 
  = \frac{\zeta_1}{\sqrt{T}} + \frac{3}{\sqrt{T}} \sum_{t=1}^T \left(1-\frac tT\right) g_L(t) \ ,
  \label{eq:skew}
\end{equation}
where $\zeta_1$ is the skewness of daily returns and 
$g_L(t)$ is the \emph{leverage correlation function} of daily returns $g_L(t) = \langle r_i r^2_{i+t} \rangle_c /\sigma^{3}$, 
which was studied in, e.g. \cite{leverage,Perello}. For IID returns, 
$g_L(t) \equiv 0$ and the skewness decays as $T^{-1/2}$. Negative price-volatility correlations produce anomalous skewness, 
that can even {\it grow} with maturity before decaying to zero.  An example of this is shown in Fig.~1 for 
a collection of international indices. A good fit of $g_L(t)$ can be obtained with a pure exponential: $g_L(t)=-A \exp(-t/t_L)$, as also
shown in Fig. 1 for the OEX. This functional form for the leverage correlation leads to the following explicit shape for the skewness:
\begin{equation}
  \zeta(T) 
  = \frac{\zeta_1}{\sqrt{T}} -
  \frac{3A}{T^{3/2}}\left(Tt_L - t_L^2(1-\exp(-T/t_L))\right) \ .
  \label{eq:skew2}
\end{equation}
It is easy to check that the leverage induced term first increases as $T^{1/2}$, reaches a maximum for $T \approx 2 t_L$ 
and decays back to zero as $T^{-1/2}$ for large maturities. 

\begin{figure}[]
  \label{fig:skewidx}
  \includegraphics[angle=-90,width= 0.7\textwidth]{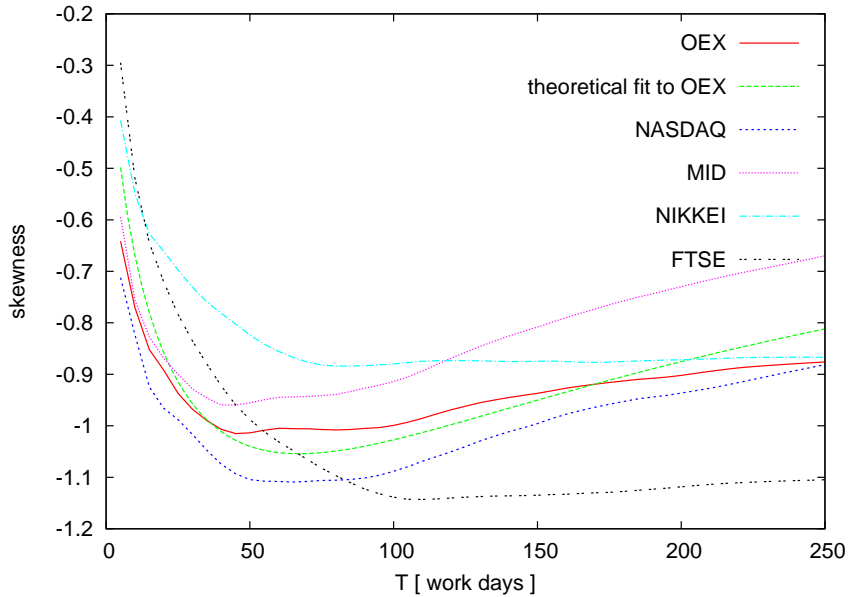}
  \caption{Historical skewness of some of the major indexes as a function of the horizon $T$, 
  computed from Eq.~\ref{eq:skew}. The data used refer to the period
  1990-2006. The SPX skewness (not shown) is nearly indistinguishable from that of the OEX. 
  The actual shape of this curve is however found to vary quite significantly
  with the time period. We also show the fit for the OEX with Eq.~\ref{eq:skew2}, leading to $A=0.16$ and $t_L=31$ days.}
\end{figure}

It is also interesting to measure the historical skewness of individual stocks,
which is much less pronounced than the index leverage \cite{leverage} -- see Figs. 2-a,b,c for small, mid and large caps, and Fig. 2-d
for a direct comparison between different market caps. Here again 
a purely exponential fit is acceptable, with however parameters $A$ and $t_L$ that depend on 
the stock, mostly through market capitalisation $M$. For the period 2001 -- 2006, we find that $t_L \sim 12$ days across all $M$s, 
whereas $A$ increases by a factor $2$ to $3$ between $M= 5\, 10^8 $ and $M= 5\, 10^{10} \$$.
A possible intuitive explanation for this increase of $A$ is that the influence of the market mode is stronger 
on large cap stocks than on small cap stocks, for which the idiosyncratic contribution is larger. 
In this case, it would indeed be expected that the leverage of large cap stocks is more akin to an index leverage effect.

\begin{figure}[]
  \label{fig:skew3}
  \includegraphics[angle=-90,width= 0.42\textwidth]{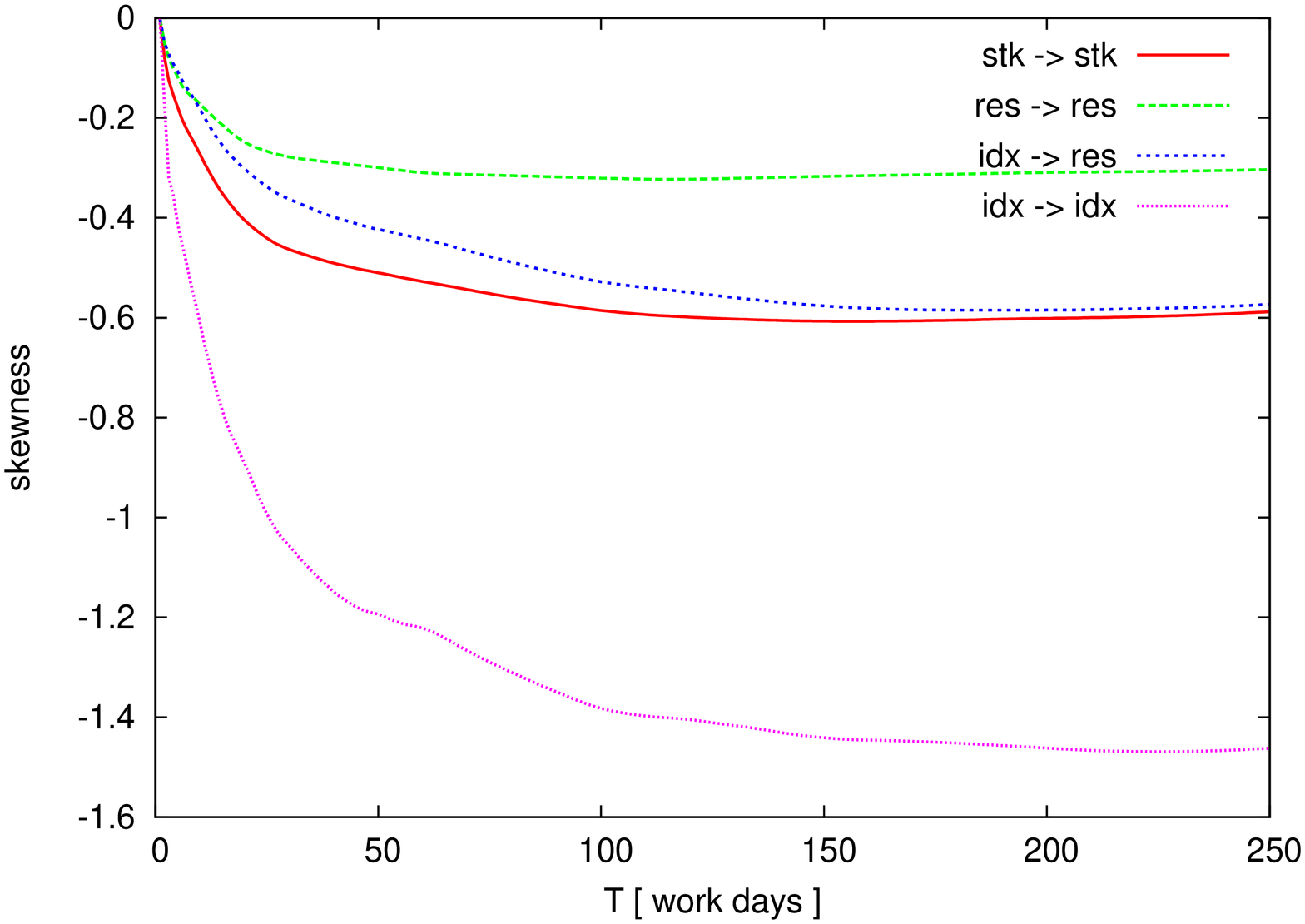}
  \includegraphics[angle=-90,width= 0.42\textwidth]{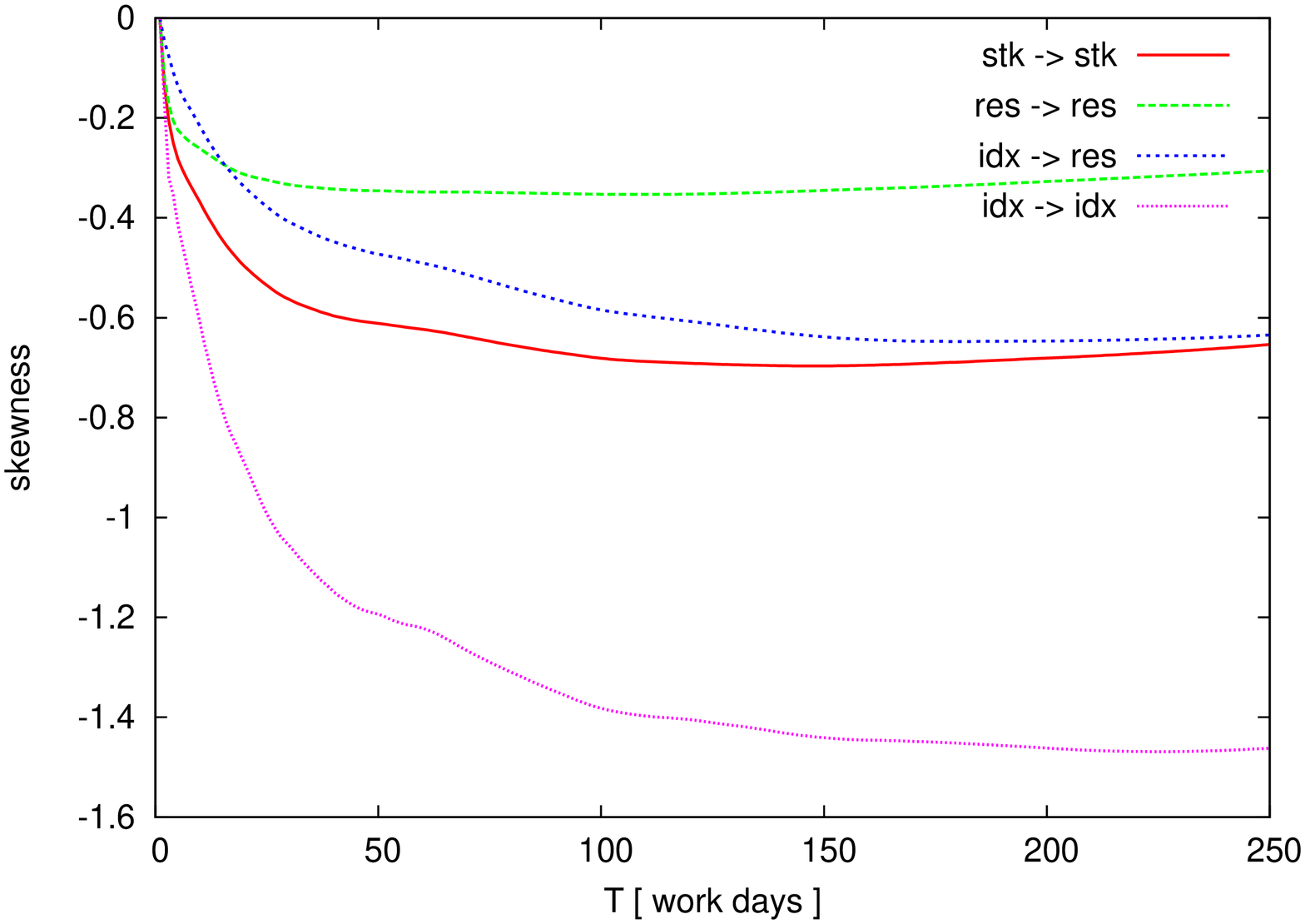}
  \includegraphics[angle=-90,width= 0.42\textwidth]{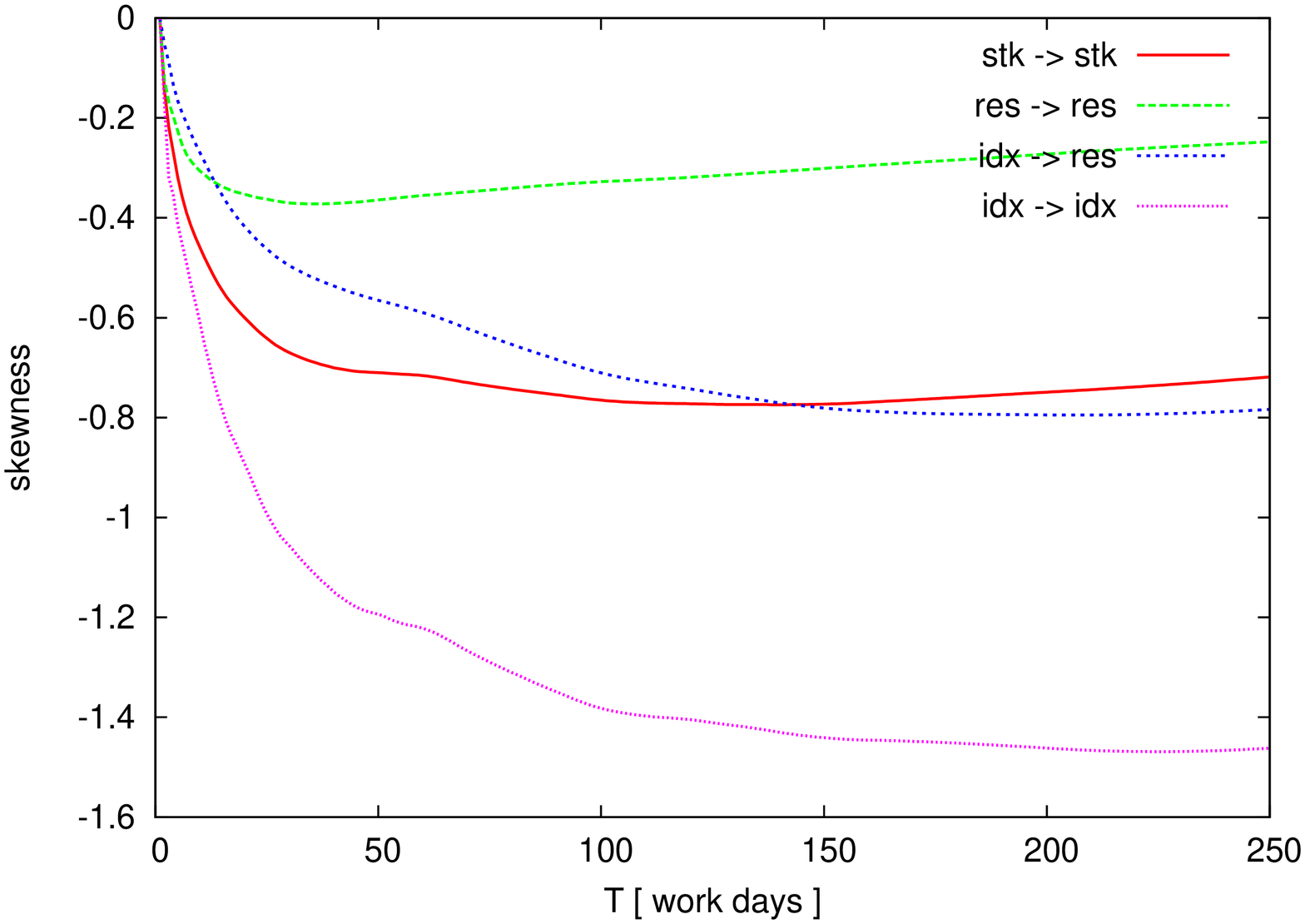}
  \includegraphics[angle=-90,width= 0.42\textwidth]{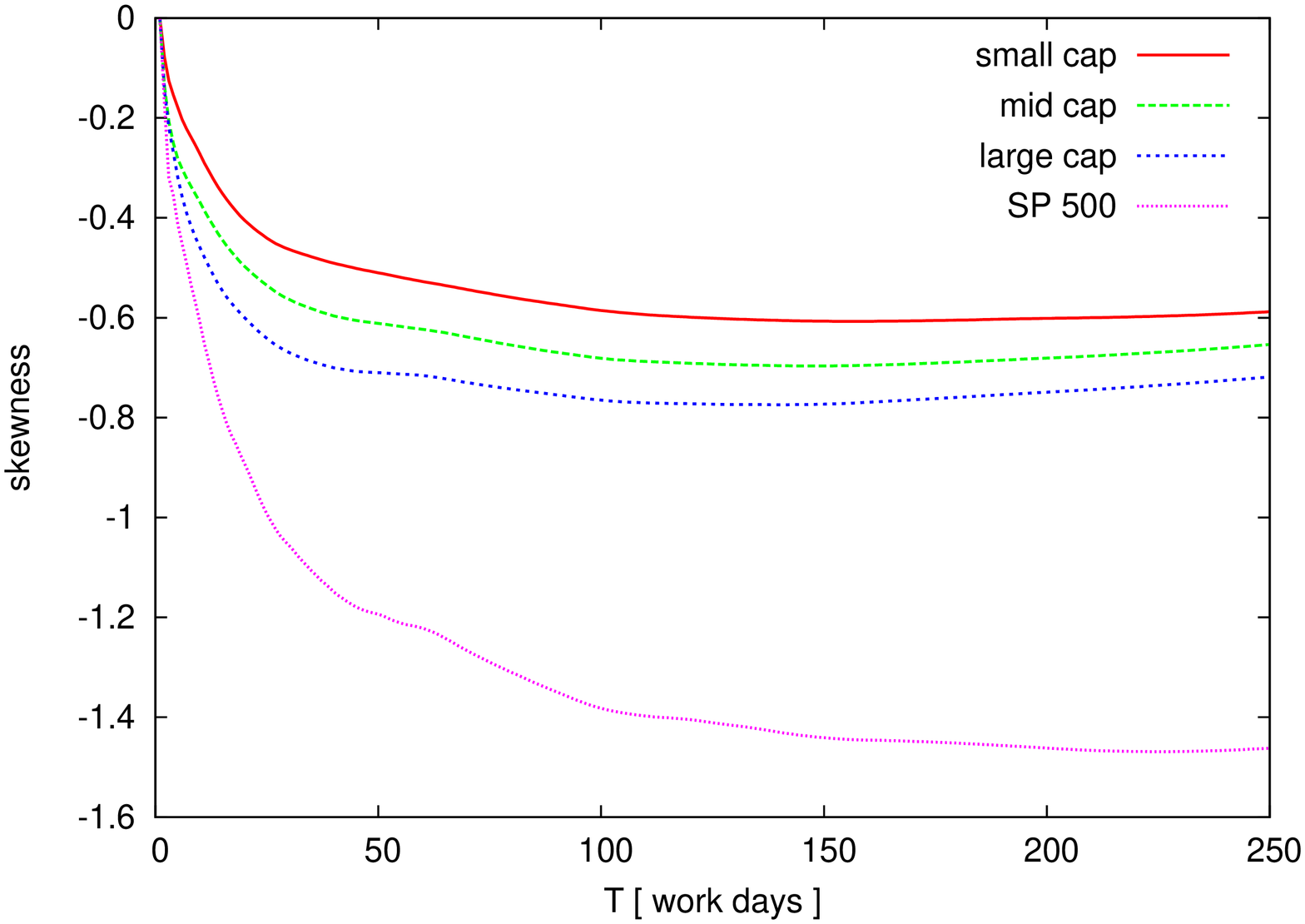}
  \caption{Historical skewness for individual stocks (symbols with error bars) and decomposition using a one-factor model, Eq.~(\ref{eq:zeta3}). 
  Top left, top right, and bottom left panels refer to (resp) small, mid, and large cap US stocks, in the period 2002-2008. In each panel, 
  we show the three
  contributions appearing in Eq.~\ref{eq:zeta3}, together with the total stock skewness. The notable feature is that the index/idiosyncratic 
  contribution $\zeta^{\Phi\to\varepsilon}(T)$ 
  increases with market cap. The bottom right compares the skewness of different market capitalisations and the index skewness.}
\end{figure}

In order to better understand this phenomenon, we postulate a simple
one-factor model for the returns of a given stock: $r_t = \beta \Phi_t +
\varepsilon_t$ , where $\Phi$ is the market return (which we approximate by the S\&P500), and
$\varepsilon_t$ is the idiosyncratic term, uncorrelated with $\Phi_t$.  
The total volatility $\sigma^2$ is decomposed into $\sigma^2=\beta^2 \sigma_\Phi^2 + \sigma_\varepsilon^2$; 
typically the second term is two to three times larger than the first one. 
The total skewness of a stock can be also be decomposed into 
three different contributions: market-induced vol on the market mode, market-induced vol on the idiosyncratic
part and idiosyncratic-induced vol on itself. More precisely, one can write:\footnote{There is in principle a 
fourth contribution describing the effect of the idiosyncratic part of the return of future market volatility, 
but is is found to be extremely small, as expected intuitively.}
\begin{equation}
  \zeta(T) =  \zeta^{\varepsilon\to\varepsilon}(T)  +
  \left(\frac{\beta \sigma_\Phi}{\sigma}\right) \zeta^{\Phi\to\varepsilon}(T) + 
  \left(\frac{\beta \sigma_\Phi}{\sigma}\right)^3 \zeta^{\Phi\to\Phi}(T) \ ,
  \label{eq:zeta3}
\end{equation}
with obvious notations.
Note that the market-induced vol contributions are weighted by the ratio of the market factor vol to the total
volatility, which decreases for smaller cap stocks. The three $\zeta$'s are all of the same order of magnitude, with 
$\zeta^{\Phi\to\Phi}(T)$ a factor 2-3 larger than the other two. Interestingly the other two contributions cross as a function of
$T$: the idiosyncratic-induced vol on itself is dominant at small $T$, whereas the market-
and their maturity dependence is shown in Fig.~2 for small, mid and large cap stocks, together with
the average total leverage effect. We have checked that the weighted contribution of the three terms add up 
the total effect, as should be. In Fig. 3 we also show the ratio $\beta \sigma_\Phi /\sigma$ as a function of the market cap 
$M$. Although this ratio indeed increases with $M$, another effect that explains the increase of the leverage
effect with market cap is the growth of the index/idiosyncratic leverage effect $\zeta^{\Phi\to\varepsilon}(T)$ (see Fig.~2).

\begin{figure}[]
  \label{fig:betaovervol}
  \includegraphics[angle=-90,width= 0.7\textwidth]{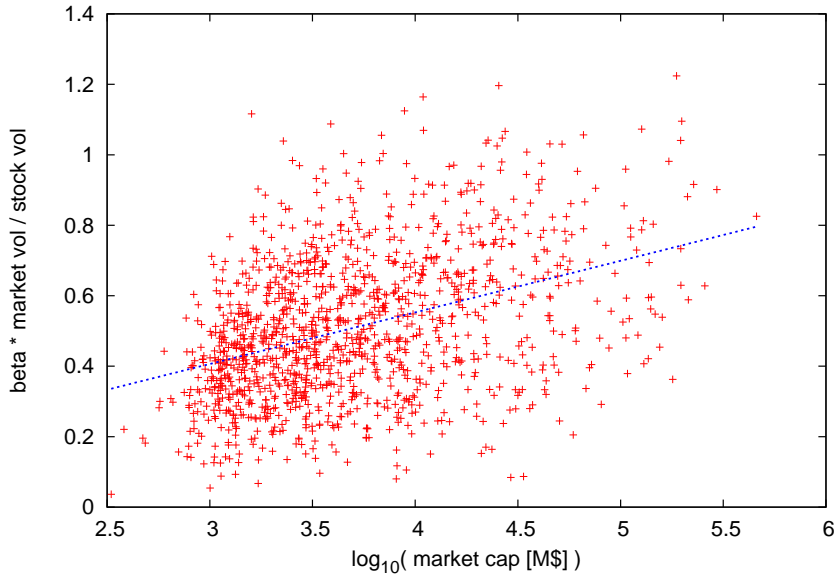}
  \caption{Ratio $\beta \sigma_\Phi/{\sigma}$ as a function of market cap. When raised to the power three as in Eq.~\ref{eq:zeta3}, the 
  market leverage effect contributes to roughly a fourth of the total skewness $\zeta(T)$.}
\end{figure}

\section{The Implied Leverage Effect}

We now turn to the dynamics of the smile, in particular of the {\it implied} leverage effect that  
measures how the at-the-money (ATM) implied volatility is correlated with the stock return. For a given day $t$, the
implied vol reads, to first order in moneyness: 
\begin{equation}\label{volexp}
\Sigma_t({\cal M},T) \approx \sigma_t(T) \left[ 1 + \frac{\zeta(T)}{6} {\cal M}  \right]
\end{equation}
where
\begin{equation}
  \sigma_t^2(T) = E[\frac 1T \int_t^{t+T} r^2_{t'} dt']
\end{equation}
is the expected average squared volatility between now and maturity, and is equal, within this approximation, to the ATM vol:
\begin{equation}
  \Sigma_t({\cal M}=0,T)=\sigma_t(T)
\end{equation}

Now, between $t$ and $t+1$, the price evolves as $S_t \to S_t + r_t S_t$. How is the smile expected to react? 
There are two simple rules of thumb commonly used in the market place:
\begin{itemize}
\item 
  {\bf Sticky Strike (ss):} The implied volatility of an option is only a function of the strike, but does not depend 
  on the price of the underlying (at least locally). Formally, $\Sigma_{t+1}(K|S_{t+1}) = \Sigma_t(K|S_t)$. 
  From the above general formula, the change of volatility should be proportional to:
  \begin{equation}
  \frac{\partial \Sigma_t}{\partial S_t} = \frac{\partial \sigma_t(T)}{\partial S_t} \left[ 1 + \frac{\zeta(T)}{6} {\cal M}  \right]
  + \frac{\sigma_t(T)\zeta(T)}{6}\frac{\partial {\cal M}}{\partial S_t} 
  \end{equation}
  Setting this derivative to zero and focusing to the ATM vol (${\cal M}=0$), one deduces:
  \begin{equation}
    \partial \Sigma_t(0,T) |_{ss} \approx \frac{\zeta(T)}{6S \sqrt{T}} \partial S
  \end{equation}
\item 
  {\bf Sticky Delta (s$\Delta$):} In this case, the smile is assumed to move with the underlying, so that the implied
  volatility of a given moneyness does not change. In particular, the ATM vol does not change:
\begin{equation}
    \partial \Sigma_t(0,T) |_{s\Delta} = 0 
  \end{equation}
\end{itemize}

A purely historical theory of the implied volatility leads to a prediction that is in between the above rules of thumb.
One starts from the above implied vol expansion Eq. (\ref{volexp}) and consider the impact of the
change in the price both on the expected future realized vol and on the moneyness. More precisely, the ATM vol evolves as:
\begin{equation}
\partial\ln \Sigma_t(0,T) = \partial \ln \sigma_t(T) - \frac{\zeta(T)}{6\sigma_t\sqrt{T}} \partial \ln S.
\end{equation}
From the definition of the leverage correlation function, and neglecting higher order (kurtosis) correlations, the 
expected relative change of the future realized volatility is given by:
\begin{equation}
\delta \sigma_t(T) = \left[\frac{1}{2T} \int_{0}^T du g_L(u)\right] r_t 
\end{equation}
Collecting all contributions, one finds that the change of ATM vol for a given stock return $r$ and a given maturity $T$
reads (we drop the $t$ dependence):
\begin{equation}
\frac{\delta \Sigma(0,T)}{\Sigma(0,T)} = \frac {1}{2\sigma(T) T}\left[\frac {1}{T} \int_{0}^T du\ u \ g_L(u) + \frac{\zeta_1}{3}
\right] 
r  \equiv \gamma(T) r,  
\end{equation}
where we have defined the {\it implied leverage} coefficient $\gamma(T)$. We assume, as above, an exponential behaviour for the 
leverage correlation function, $g_L(u)= -A  \exp(-t/t_L)$ and neglect the $\zeta_1$ contribution (see below). We then find that the theoretical 
implied leverage is given by:
\begin{equation}
\gamma(T)|_{th}= -\alpha  \left( \frac {1 - (1+\tilde T) e^{-\tilde T}} {\tilde
  T^2} \right);\quad \tilde T \equiv T/t_L \, 
\end{equation}
with $\alpha \equiv {A}/{2\sigma(0)}$.
This is the central result of this study. It should be compared to ones obtained from the sticky strike/sticky delta prescriptions:
\begin{equation}
  \gamma(T)|_{ss} \approx -\alpha 
  \left( \frac 1 {\tilde T}  - \frac{(1-e^{-\tilde T})} {\tilde T^2} \right); \qquad \gamma(T)|_{s\Delta} = 0.
\end{equation}
The asymptotic behaviour of these quantities can be compared, one finds $\gamma(T \to 0)|_{th}=\gamma(T \to 0)|_{ss}=-A/{4\sigma(0)}$
(in agreement with the general result of Durrlemann \cite{Durrlemann}) and $|\gamma(T \to \infty)|_{th}| \propto T^{-2} \ll 
|\gamma(T \to \infty)|_{ss}|  \propto T^{-1}$. In other words, the sticky strike always overestimates 
the implied leverage, but becomes exact for short maturities compared to the leverage correlation time: $T \ll t_L$. 
The sticky delta procedure, on the other hand, always underestimates the true implied leverage, but becomes 
a better approximation than the sticky strike for large maturities.

\section{Comparison with empirical data: indices and individual stocks}

These predictions are compared with empirical data on the implied leverage effect on the OEX index, large cap, mid cap and small cap stocks, 
in Figs. 4-7. On these plots we show the implied leverage coefficient $\gamma(T)$ as a function of maturity. The three curves correspond to (a) the theoretical 
prediction $\gamma(T)|_{th}$ computed using the historically determined leverage correlation $g_L(t)$, (b) the sticky strike procedure 
$\gamma(T)|_{ss}$ using the same historical parameters and (c) the data obtained using from the daily change of ATM implied volatilities,
$\gamma(T)|_{imp}$. We also show the $\gamma = 0$ line corresponding to sticky delta. The implied data is obtained by regressing the relative
daily change of ATM implied vols on the corresponding stock or index return, for each maturity. The result is then averaged over all stocks within
a given tranche of market capitalisation. Similarly, the coefficient $\alpha$ needed to compute the theoretical prediction is obtained
by fitting the leverage correlation for each stock individually, normalizing it by the realized volatility over the same period, and then 
averaging the ratio $A/\sigma_t$ across different stocks. It turns out that $A$ itself is to a good approximation proportional to $\sigma_t$ anyway \cite{leverage}.

It is clear from these plots that on average, implied volatilities overreact to
changes of prices compared to the prediction calibrated on the historical leverage effect, except maybe for small cap stocks where the level of $\gamma$
is in the right range at short maturities. The overestimation tends to grow with maturity, since the theoretical prediction is that $\gamma(T \to \infty)|_{th} 
\sim T^{-2}$ whereas the implied value $\gamma(T)|_{imp}$ appears to saturate at large $T$. In fact, the $\gamma(T)|_{imp}$ curve 
appears to be well fitted by a sticky-strike prediction $\gamma(T)|_{ss}$, but with an effective value of the parameter $A$ substantially
larger than its historical value. This would be compatible with the fact that market makers use a simple sticky strike procedure, but with a smile
that is significantly more skewed than justified by historical data. We only have partial evidence that the implied skew is indeed too large, but
cannot check this directly with the data at our disposal at present. However, we believe that this is a very plausible explanation to our findings. We have
furthermore checked that our conclusions are stable over different time periods.

In the above theory for $\gamma$, we have explicitely neglected the one-day skewness $\zeta_1$. Could this term explain the above
discrepancy? We have checked that this term gives a small contribution to $\gamma$ -- at most $10 \%$ for short maturities. In fact, 
the daily skewness of individual stocks is even slightly positive, which should lead to a further (small) reduction of the implied 
skew and of the implied leverage effect. 

Finally, we have studied more systematically the dependence of $\gamma(T)|_{imp}$ on market capitalization $M$. The results are 
shown in Fig. 8. As expected (and already clear from Figs. 4-6), $\gamma(T)|_{imp}$ increases (in absolute
value) with $M$; a good fit of the dependence
is:
\begin{equation}
  \gamma(T)|_{imp} \approx a(T) + b(T) \ln M 
\end{equation}
where $a$ and $b$ are maturity dependent coefficients.

\section{Conclusion}

In this paper, we have provided a first principle theory for the implied volatility skew and at-the-money implied leverage 
effect in terms of the historical leverage correlation, i.e. the lagged correlation between past returns and 
future squared returns. We have compared the theoretical prediction for the implied leverage to two well
known rules of thumb to manage the smile dynamics: sticky strike or sticky delta. The sticky strike is 
exact for small maturities, but is an upper bound to the theoretical result otherwise. The sticky delta is 
a (trivial) lower bound, which however becomes more accurate than sticky strike at large maturities. 

We have then compared these theoretical results to data coming from option markets. We find that the implied 
volatility strongly over-reacts, on average, to change of prices, especially for long dated options. A plausible
interpretation is that market makers tend to use a sticky strike rule, with an exaggerated skewness of the 
volatility smile. It would be interesting to test this hypothesis directly, with full option smile data and 
not only at-the-money vols like in this study. 

We have also provided an empirical study of the market cap dependence of these effects. We find that both the
historical and implied leverage effect is stronger for larger cap stocks, with a roughly logarithmic dependence
on market cap. Although this is partly explained 
in terms of the ratio of the idiosyncratic vol to the total vol (smaller for larger cap stocks), we 
find that the index/idiosyncratic leverage effect is stronger for these large caps. This may 
relate to the risk aversion interpretation of the leverage effect put forth in \cite{leverage}: since large cap
stocks are followed by more market participants, feedback effects could be stronger there.

\begin{figure}[]
  \label{fig:gammalc}
  \includegraphics[angle=-90,width= 0.7\textwidth]{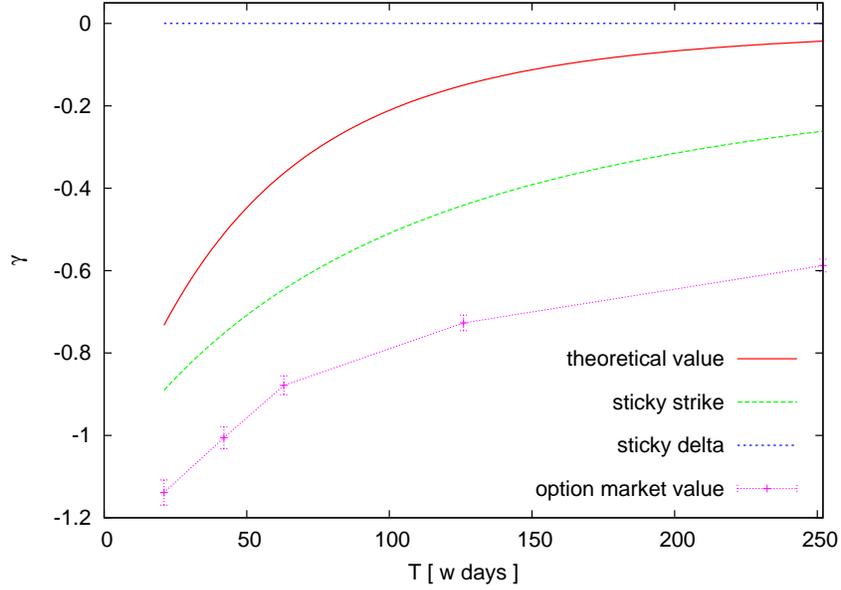}
  \caption{Comparison between the implied leverage coefficient $\gamma(T)|_{imp}$ and the various theoretical predictions, 
  sticky strike $\gamma(T)|_{ss}$, striky delta $\gamma(T)|_{s\Delta}=0$, and historical $\gamma(T)|_{th}$, for 
  large cap US stocks in the period 2004-2008. Note that the implied volatility over-reacts to stock moves, in particular for 
  large maturities. The effect can be roughly accounted for by the use of a sticky-strike rule with an implied skewness overestimated 
  by $\sim 50 \%$ at short maturities and $\sim 100\%$ at large maturities. The error bars are statistical and assume all stocks in the
  pool behave the same way.
  }
\end{figure}

\begin{figure}[]
  \label{fig:gammamc}
  \includegraphics[angle=-90,width= 0.7\textwidth]{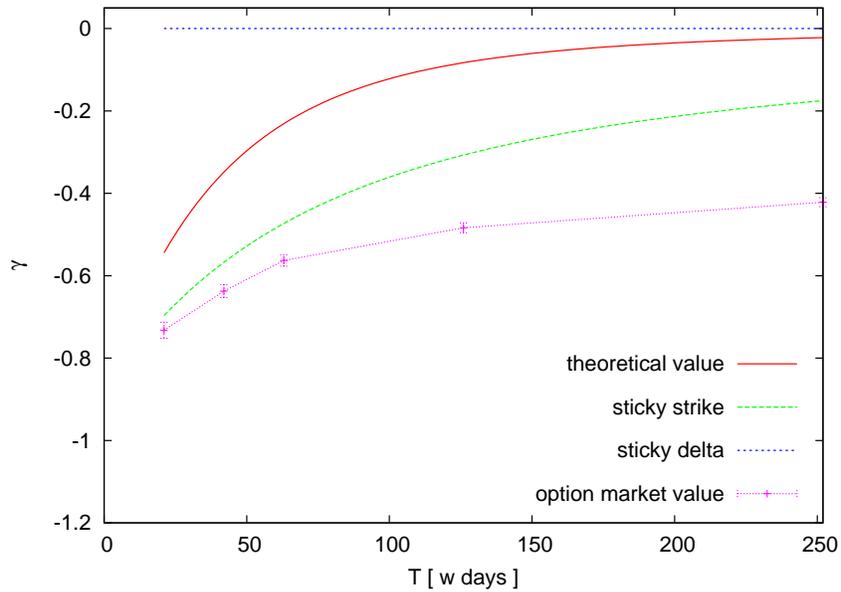}
  \caption{Same as Fig. 4, but for mid-cap US stocks.}
\end{figure}

\begin{figure}[]
  \label{fig:gammasc}
  \includegraphics[angle=-90,width= 0.7\textwidth]{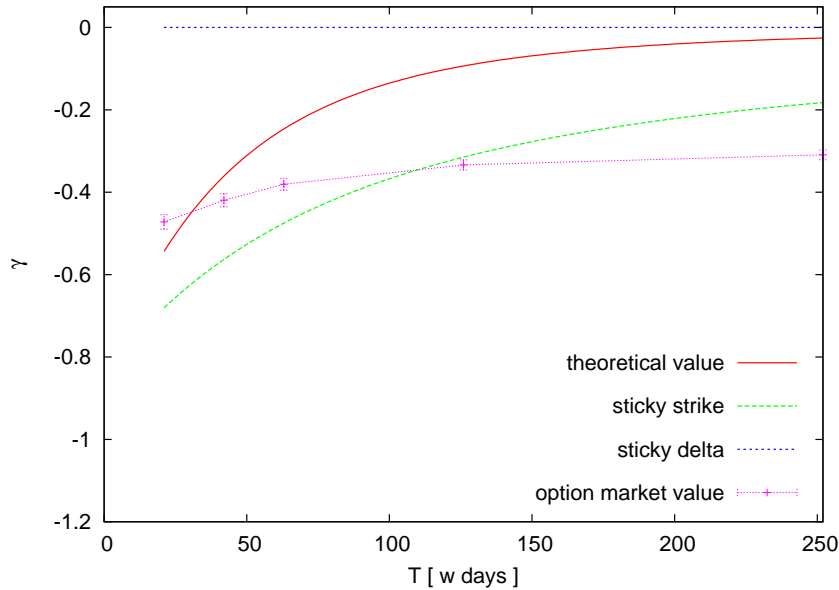}
  \caption{Same as Fig. 4, but for small-cap US stocks. In this case, the implied leverage appears to have the correct amplitude
  for small maturities, but the term structure of $\gamma(T)$ is anomalously flat here.}
\end{figure}

\begin{figure}[]
  \label{fig:oex}
  \includegraphics[angle=-90,width= 0.7\textwidth]{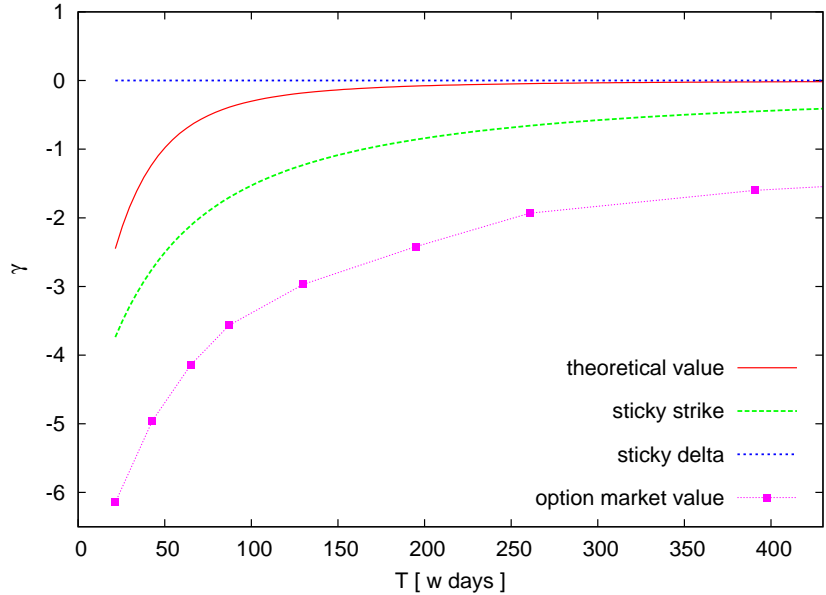}
  \caption{Same as Fig. 4, but for the OEX index. Now the implied skewness needed to explain the data with sticky strike 
  is overestimated by a factor $2$ to $3$.}
\end{figure}

\begin{figure}[]
  \label{fig:gammavsmcap}
  \includegraphics[angle=-90,width= 0.7\textwidth]{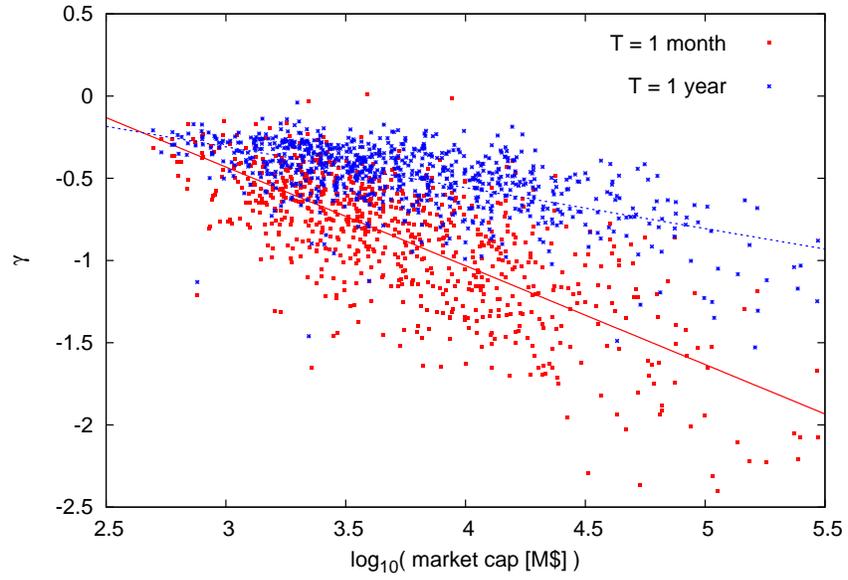}
  \caption{Scatter plot of the implied leverage coefficient $\gamma(T)|_{imp}$ as a function of the log market cap, and linear regression
  for $T=1$ months and $T=1$ year. The linear regressions are, respectively: $\gamma(T)|_{imp}=1.37 - 0.60 \log_{10} M $ and
  $\gamma(T)|_{imp}=0.44 - 0.25 \log_{10} M $.}
\end{figure}

\end{document}